\documentclass{ws-procs975x65}

\newcommand{\lanln}[1]{$\langle$\texttt{arXiv:#1}$\rangle$}

\begin{document}

\title{A NEW EXPRESSION FOR THE TRANSITION RATE OF AN ACCELERATED PARTICLE DETECTOR\footnote{This research was supported by
an EPSRC Dorothy Hodgkin Research Award to the University of Nottingham}}

\author{JORMA LOUKO\footnote{jorma.louko@nottingham.ac.uk} and ALEJANDRO SATZ\footnote{pmxas3@nottingham.ac.uk}}

\address{School of Mathematical Sciences,
University of Nottingham,\\
 Nottingham NG7 2RD, UK}

\begin{abstract}

We analyse the instantaneous transition rate of an accelerated Unruh-DeWitt particle detector whose coupling to a quantum field on Minkowski space is regularised by a finite spatial profile. We show, under mild technical assumptions, that the zero size limit of the detector response is well defined, independent of the choice of the profile function, and given by a manifestly finite integral formula that no longer involves epsilon-regulators or limits. Applications to specific trajectories are discussed, recovering in particular the thermal result for uniform acceleration. Extensions of the model to de Sitter space are also considered. 

\end{abstract}

\bodymatter

\section{Introduction}

The Unruh-DeWitt particle detector model \cite{unruh,deWitt} is a useful tool for probing the physics of quantum fields. The simplest case to consider is an idealised two-state atom with a monopole coupling to a massless scalar field in its Minksowski vacuum state. Up to a detector-dependent proportionality constant, the probability of a transition of energy $\omega$ at proper time $\tau$, after ``turning on'' the interaction at proper time $\tau_0$, is given in first-order perturbation theory by the response function
\begin{equation}
 \label{defresponse-tau}
F_\tau(\omega)= 
\int_{\tau_0}^{\tau}
\mathrm{d}\tau' \int_{\tau_0}^{\tau}\mathrm{d}\tau'' 
\, 
\mathrm{e}^{-i\omega(\tau'-\tau'')}
\, 
W(\tau',\tau'')\,,
\end{equation}
where $W(\tau',\tau'')=\langle 0\vert \phi(\mathsf{x}(\tau'))\phi(\mathsf{x}(\tau''))\vert 0\rangle $ is the Wightman function of the field. The $\tau$-derivative of the response function is the transition rate
\begin{equation}
\label{defexcitation-sharp}
\dot{F}_{\tau} (\omega)
=
2 \, 
\mathrm{Re}
\int_{0}^{\tau - \tau_0}
\mathrm{d}s
\,\,
\mathrm{e}^{-i\omega s}
\, 
W(\tau,\tau-s)
\, . 
\end{equation} 
Using the conventional $i\epsilon$ regularisation prescription, the Wightman function reads
\begin{equation}
\label{tradWightman}
W(\mathsf{x},\mathsf{x'})
=
\lim_{\epsilon \to 0_+} 
\frac{-1}{4\pi^2}\frac{1}{{(t-t'-i\epsilon)}^2
- 
{\vert\mathbf{x}-\mathbf{x'}\vert}^2}
\ ,
\end{equation}
with the limit being taken after integration against smooth functions of $\mathsf{x}(\tau')$ and $\mathsf{x}(\tau'')$. However, the sharp cutoff assumed in the integrals (\ref{defresponse-tau}) and (\ref{defexcitation-sharp}) implies that this form of the two-point function is not guaranteed to give unambiguous results. In fact, Schlicht \cite{schlicht} and Langlois \cite{Langlois} have shown that this procedure gives Lorentz non-invariant results for a uniformly accelerated trajectory, instead of the thermal spectrum expected according to the Unruh effect.

Schlicht \cite{schlicht} has proposed a new regularisation scheme that avoids this problem. The detector is coupled to a spatially smeared version of the field operator given by
\begin{equation}
\label{smeared}
\phi_f(\tau)  =
\int \mathrm{d}^3 \xi
\,
f_\epsilon (\boldsymbol{\xi})
\, 
\phi 
\bigl( \mathsf{x}(\tau,\boldsymbol{\xi}) \bigr)
\ , 
\end{equation}
where $f_\epsilon (\boldsymbol{\xi})$ is a profile function and $\boldsymbol{\xi}$ are Fermi-Walker coordinates parametrizing the simultaneity plane of the detector at time $\tau$. The parameter $\epsilon$ controls the size of the detector, recovering the pointlike coupling in the $\epsilon\rightarrow 0$ limit. Using a particular Lorentzian profile function, Schlicht obtained the modified correlation function
\begin{equation}
\label{corrSchlicht}
W_\epsilon (\tau,\tau')
=\lim_{\epsilon \to 0_+}
\frac{1}{4\pi^2}\frac{1}{\bigl( \mathsf{x}-\mathsf{x}'
-i\epsilon 
(\dot{\mathsf{x}}+\dot{\mathsf{x}}') 
\bigr)^2}
\ , 
\end{equation}
and showed that using it in (\ref{defexcitation-sharp}) gives the correct Planckian result for the Rindler motion. But it remained open whether this result depends on the choice of a convenient profile. This is a motivating question for our research.

\section{Results}

Our first result is that it is possible to take the explicit $\epsilon\rightarrow 0$ limit in Schlicht's expression for the transition rate, with the outcome
\begin{equation}
\label{resultado1}
\dot{F}_{\tau}(\omega)
=
-\frac{\omega}{4\pi}+\frac{1}{2\pi^2}
\int_0^{\Delta\tau}\textrm{d}s
\left( 
\frac{\cos (\omega s)}{{(\Delta \mathsf{x})}^2} 
+ 
\frac{1}{s^2} 
\right) 
\ \ +\frac{1}{2\pi^2 \Delta \tau}
\ ,
\end{equation}
where ${(\Delta \mathsf{x})}^2=(\mathsf{x}(\tau)-\mathsf{x}(\tau-s))^2$ and $\Delta\tau=\tau-\tau_0$. 

Formula (\ref{resultado1}) contains no regulators and is manifestly Lorentz invariant. It separates the spectrum cleanly into a universal term odd in $\omega$ and a trajectory-dependent term even in $\omega$. It is therefore a convenient starting point for concrete calculations of detector response for generic trajectories (the only condition imposed on $\mathsf{x}(\tau)$ to derive the result is $C^9$ continuity). Asuming some further but still mild conditions on the trajectory to control the asymptotic past limit, expression (\ref{resultado1}) is also valid when $\Delta\tau=+\infty$. All stationary motions in Minkowski space are covered under these assumptions, as well as many nonstationary ones (excluded are certain pathological cases like trajectories that cover an infinite space in a finite proper time). In particular, for the Rindler motion our formula obtains the Planckian spectrum, and for certain asymptotically Rindler motions an asymptotically Planckian spectrum. For the general case with $\Delta\tau=+\infty$ the spectrum can also be written as
\begin{equation}
{\dot F}_{\tau}(\omega)
= 
-\frac{\omega}{2\pi}\Theta(-\omega)
+\frac{1}{2\pi^2}\int_0^{\infty}\textrm{d}s\cos (\omega s)
\left( 
\frac{1}{{(\Delta \mathsf{x})}^2} 
+\frac{1}{s^2}
\right) 
\ , 
\label{inermasacc}
\end{equation}
where the first term is the spectrum for inertial motion and the second contains the effects of acceleration.

Our second result is that the same expression (\ref{resultado1}) also follows in the zero-size limit from \textit{any} profile function $f_\epsilon (\boldsymbol{\xi})$ which has compact support, if a technical modification is made to the definition of spatial smearing so that the transition rate is defined by
\begin{equation}
\label{eq:smearedFdot-def}
\dot{F}_{\tau}^{\epsilon}(\omega)
=
\int_{\boldsymbol{\xi} \ne \boldsymbol{\xi}'}
\mathrm{d^3}\xi
\,
\mathrm{d^3}\xi' 
\; 
f_{\epsilon}(\boldsymbol{\xi}) 
\, 
f_{\epsilon}(\boldsymbol{\xi}')
\, 
2 \,
\mathrm{Re}\int_{0}^{\Delta\tau} \mathrm{d}s
\,\,
\mathrm{e}^{-i \omega s}
\, 
\langle 0\vert 
\phi \bigl(\mathsf{x} (\tau,\boldsymbol{\xi}) \bigr)
\phi \bigl(\mathsf{x}(\tau-s,\boldsymbol{\xi}') \bigr)
\vert 0\rangle
\ . 
\end{equation}
The $\epsilon\rightarrow 0$ limit of (\ref{eq:smearedFdot-def}) can be taken in a general way to obtain (\ref{resultado1}), assuming the trajectory to be real analytic. Full details leading to these two main results can be found in our paper \cite{us}.

Thirdly, the second result can be easily generalised to de Sitter spacetime (dS). For a detector moving in dS when the field is in the Euclidean vacuum state, the zero-size limit of the transition rate for a general compact profile is given by
\begin{equation}
\label{resultadoDS}
\dot{F}_{\tau}(\omega)
=
-\frac{\omega}{4\pi}+\frac{1}{2\pi^2}
\int_0^{\Delta\tau}\textrm{d}s
\left( 
\frac{\cos (\omega s)}{\left[ \mathsf{Z}(x(\tau))-\mathsf{Z}(x(\tau-s))\right]^2 } 
+ 
\frac{1}{s^2} 
\right) 
\ \ +\frac{1}{2\pi^2 \Delta \tau}
\ ,
\end{equation}
where $\mathsf{Z}(x)$ are the Minkowski coordinates corresponding to de Sitter point $x$ in a five-dimensional space in which dS is embedded as an hyperboloid $\mathsf{Z}^2=\alpha^2$. Expression (\ref{resultadoDS}) is manifestly de Sitter invariant and gives the expected Planckian spectrum for inertial trajectories.

\section{Conclusions and outlook}

We have calculated the zero-size limit of the transition rate for particle detectors regularised by a spatial profile. The result, given by expression (\ref{resultado1}), was applied to a number of trajectories in Minkowski space and generalises straightforwardly to de Sitter space. Whether similar expressions hold in more general backgrounds is an open question. It is worth remarking that when the conventional regularisation is used and the detector is switched on and off with a smooth function, expression (\ref{resultado1}) is also recovered as the approximate transition rate in the fast switching limit \cite{us2}. This suggests its universal status.

\vfill

\end{document}